\documentclass{icrc2009}

\usepackage{graphicx}   
\usepackage[caption=false]{caption}    
\usepackage[font=footnotesize]{subfig} 
\usepackage{fixltx2e}
\usepackage{url}

\newcommand{\shorttitle}[1]%
{\markboth{Proceedings of the 31\MakeLowercase{$^{st}$} ICRC, {\L}\'{o}d\'{z} 2009}{#1} }  
\newcommand{\etal}{\MakeLowercase{\textit{et al. }}} 

\begin{document}
\title{Performance of the Camera of the MAGIC II Telescope}

\author{\IEEEauthorblockN{D. Borla Tridon\IEEEauthorrefmark{1},
    F. Goebel$\dag$\IEEEauthorrefmark{1}, 
    D. Fink\IEEEauthorrefmark{1}, 
    W. Haberer\IEEEauthorrefmark{1},
    J. Hose\IEEEauthorrefmark{1},
    C.C. Hsu\IEEEauthorrefmark{1},
    T. Jogler\IEEEauthorrefmark{1},
    R. Mirzoyan\IEEEauthorrefmark{1},\\
    R. Orito\IEEEauthorrefmark{1},
    O. Reimann\IEEEauthorrefmark{1},
    P. Sawallisch\IEEEauthorrefmark{1},
    J. Schlammer\IEEEauthorrefmark{1},
    T. Schweizer\IEEEauthorrefmark{1},
    B. Steinke\IEEEauthorrefmark{1},
    M. Teshima\IEEEauthorrefmark{1} \\
    on behalf of the MAGIC Collaboration
\IEEEauthorblockA{\IEEEauthorrefmark{1}Max Planck Institute f$\ddot{u}$r Physik, F$\ddot{o}$hringer Ring 6, D-80805 Munich, Germany}
}}
\shorttitle{D. Borla Tridon \etal MAGIC II Camera}
\maketitle

\begin{abstract}
MAGIC comprises two 17m diameter IACTs to be operated in stereo mode.
Currently we are commissioning the second telescope, MAGIC II. The camera of the second telescope has been equipped with 1039 pixels of 0.1-degree diameter.
Always seven pixels are grouped in a hexagonal configuration to form a cluster.
This modular design allows easier control and maintenance of
the camera. The pixel sensors are high quantum efficiency photomultiplier tubes (PMTs) from Hamamatsu
(superbialkali type, QE $\sim32\%$ at the peak wavelength) that we operate at rather low gain of $3\cdot 10^4$. This allows us to also perform extended observations under moderate moonlight. The system of two MAGIC telescopes will at least double the sensitivity compared to MAGIC I and also will allow us to lower the energy threshold.
Here we will report the performances of the Camera of the second MAGIC telescope.
\end{abstract}

 
\section{INTRODUCTION}
MAGIC is a system of two 17 m diameter mirror imaging atmospheric Cherenkov telescopes (IACT) for very high energy, ground-based gamma ray astronomy. The first telescope, MAGIC-I, is in operation since 2004. The second telescope was completed in 2008 and is  now under commissioning. We plan to operate it within a few months. The two telescope system will be operated in stereo mode and have an improved sensitivity and a lower energy threshold. The latter will have a strong impact on pulsar studies and will extend the accessible red shift range, which is limited by the absorption of high energy $\gamma$-rays by the extragalactic background light.\\
New components of improved performance are used for the camera of MAGIC II. In particular, new photomultipliers (PMT) of increased peak quantum efficiency (QE) of around $32\%$ are used in the first phase while it is planned to use even higher QE hybrid photo detectors (HPD) in the second phase. Moreover, the entire signal chain from the PMTs to the FADCs is redesigned to have a bandwidth of 500 MHz. The Cherenkov light flashes from low energy $\gamma$-ray showers are very short in time ($\leq 2.5ns$ at the PMT). A high bandwidth signal chain and a GHz pulse digitization will therefore allow one to minimize the integration time, thus to reduce the influence of the background from the light of the night sky (LONS) and improve the gamma/hadron separation by a better timing analysis.
\section{THE CAMERA DESIGN}
The camera of the telescope is placed in the focus of the reflector, i.e. at a distance of 17 m above the mirror dish.
MAGIC-II has a camera with a field of view (FoV) of $3.5^{o}$ diameter, equipped uniformly with 1039 pixels of $0.1^{o}$ diameter grouped into 169 cluster modules. The trigger region includes 91 cluster modules covering a FoV of $2.5^{o}$ diameter. Each cluster comprises seven pixels grouped in a hexagonal configuration. The modular design allows easier control and maintenance of the camera. On the front side the pixels are equipped with Winston cone type light guides to minimize light losses due to the dead area between the PMTs and to reject large angle background light. Both the input and the output pupils of the light catchers have a hexagonal shape.
\begin{figure}[!t]
\centering
\includegraphics*[scale=0.45]{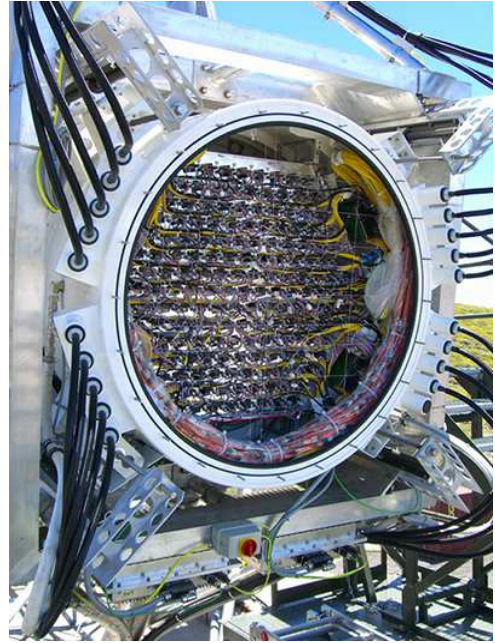}
\caption{View of the entire camera from the back side with the modular design of 169 clusters with 7 pixels each.}
\label{fig:camera}
\end{figure}

\section{CAMERA HOUSING AND COOLING SYSTEM}
Since the camera is placed in the focus of the reflector, i.e. far from the elevation axis of the telescope structure, the overall weight of camera mechanics and electronics must be minimized. Therefore most mechanical components are made from aluminum. The total weight is 600 kg for a camera dimension of 1.462 m in diameter and 0.81 m in thickness. The camera electronics is powered by two 5 V power supplies mounted in two boxes placed outside of the camera housing.\\ The central part of the camera body consists of two cooling plates in which a cooling liquid is running through pipes in order to stabilize the temperature of the camera electronics. The cooling system is designed to operate in a range of outdoor temperature from $-10$ to $+30^{o}C$; its maximum power consumption is 8kW while the total power consumption of the entire camera electronics is $\leq$ 1kW. The system has a temperature stability of $\pm 1^{o}C$.\\ The clusters are inserted from the front side into tightly machined holes of the two cooling plates thus having good thermal contact. In total 169 clusters are installed; 127 of them are fully assembled with PMTs, while 42 clusters in the circumference of the camera are only partially equipped with PMTs in order to approximate a round configuration. In addition, at six corners there is space foreseen for six more clusters, which can be used to test new photon detectors.\\ Inside the camera body on the back side of the cooling plates we installed two VME crates to which all the clusters are connected via LAN cables with RJ45 connectors. Fan-out boards, receiver boards and the slow control cluster processor (SCCP) for the starguider are also placed in the corners inside the square camera body. The camera is sealed by a 3 mm thick, UV transmitting Plexiglas window, (94$\%$ transmission above 340 nm and still of 80$\%$ transmission at 300nm).
\section{THE PMT CLUSTERS}
\begin{figure}[hbt]
\centering
\includegraphics*[scale=0.2]{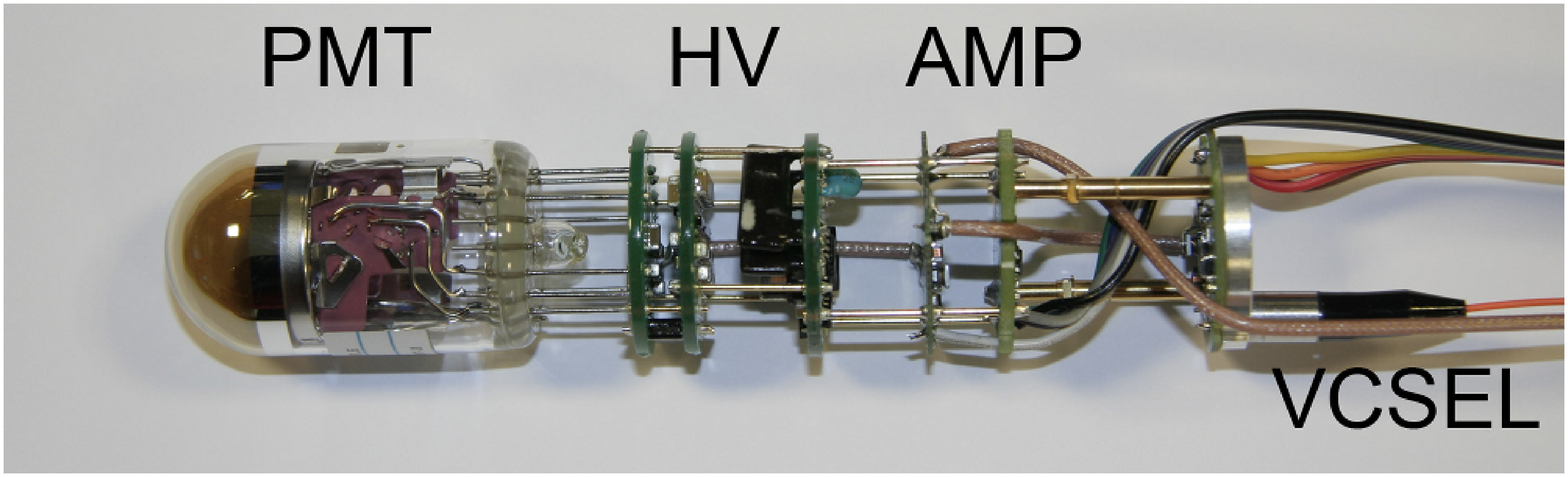}
\centering
\includegraphics*[scale=0.2]{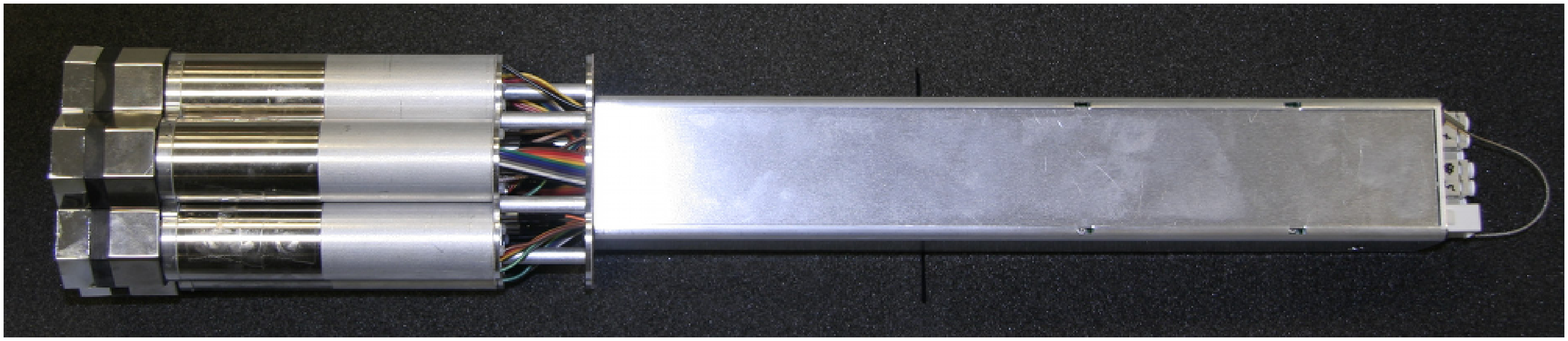}
\caption{\label{fig:pmt}
Assembled PMT module to form a pixel in the upper figure and a full cluster of 7 pixels in the bottom figure}
\end{figure}

As mentioned, the pixels are equipped with photomultiplier tubes (PMTs) from Hamamatsu (superbialkali, type R10408) of 25.4 mm diameter, a hemispherical photocathode and only 6 dynodes. Each pixel module includes a compact power unit providing the bias voltages for the PMT and a stack of round circuit boards for the front-end analog signal processing. The voltage generator was produced by Hamamatsu while the other electronics was produced at the MPI in Munich (see the configuration in the upper photo of Fig. \ref{fig:pmt}). The PMT bias voltages for the cathode and dynodes are generated by a low power, nine step Cockroft-Walton DC-DC converter, which can provide up to 1250 V peak voltage. The voltage distribution is such that the voltage between the photocathode and the first dynode is 3 times the voltage between the following dynodes. We operate the 6 dynodes PMTs at a rather low gain of typically $3\cdot10^{4}$ in order to also allow observations under moderate moonlight without damaging the dynodes.\\
The PMTs signals are amplified by 700MHz bandwidth AC coupled preamplifiers ($\sim 25dB$ amplification) placed in one of the circuit boards. The amplifier inputs are protected to guard against potentially destructive voltage spikes from the PMTs. The amplified PMT signals are back-converted to optical signals by vertical cavity laser diodes (VCSEL), which transmit the analog signals by multi-mode optical fibers of 160 m length to the readout system in the counting house.  The Avalon multimode, 850 nm emitting VCSELs are placed on the last printed board of the stack and are thermally coupled to the cooling plate in order to minimize temperature variations and thus to prevent drift of the optical power during operation. The VCSELs are biased at 3mA to achieve low mode partition noise.
 Moreover, test pulses adjustable between 0 and 1.6 V amplitude, can be injected by computer control over a 50-Ohm coax cable into the input of the amplifier. This allows functionality and linearity tests of the entire signal chain after the PMT. In addition, monitoring circuits provide readout of the average PMT current, the PMT voltage and, as well, the VCSEL monitor photodiode current and temperature.\\
The pixel modules are encapsulated by shielding aluminum tubes, which also hold the Winston cones. In addition, the PMT tubes are wrapped into Mu-metal foils to screen them from the Earth magnetic field. Always, seven pixel modules are grouped together in a cluster body. In each cluster the SCCP board for the control electronics, the power distribution and the test-pulse generator are placed inside an aluminum box as shown in the lower photo of Figure \ref{fig:pmt}.
\section{THE SLOW CONTROL OF THE CAMERA}
The camera is controlled by a number of SCCPs. A SCCP board, installed in each cluster, controls the operations of the camera and reads several parameters. The HV of each pixel is set individually by computer control and, as already mentioned, the PMT current, the HV and the temperature at the VCSEL are continuously monitored.  In addition, the slow control operates the camera lids in front of the Plexiglas window to protect the PMTs against bright light and steers the power supplies of the camera. Each SCCP has a flash programmable processor with 12 bits resolution DACs of 0-1.25 V range and 0-2.5 V range ADCs of 12 bit resolution. Each SCCP board is connected to a VME board in one of the two VME crates. The VME crates are connected to the camera control PC in the counting house via an optical PCI to VME link.
\section{THE DATA ACQUISITION SYSTEM}
The optical signals from the camera are transmitted via optical fibers to the counting house, where they are converted back to electrical signals. These are split into two branches. One branch is coupled to a discriminator with a computer controlled adjustable threshold. The discriminator signal is sent to the trigger system. The signal of the other branch is further amplified and coupled to a 2 GSample/s digitization and acquisition system based upon a low power analog sampler called Domino Ring Sampler \cite{icrc6}. The analog signals are stored in a multi capacitor bank that is organized as a ring buffer. All capacitors in the bank are sequentially enabled by a shift register driven by an internally generated 2 GHz clock locked by a phase locked loop (PLL) to a common synchronization signal. Once a trigger occurs, the acquisition of new signals is stopped and the stored signals in the ring buffer are read out at a lower frequency of 40 MHz and digitized by a 12 bit resolution ADC. Currently, the bandwidth of the entire signal chain from the PMTs to the FADCs is limited to less than 200 MHz due to the use of early development, relatively slow DOMINO-2 chips resulting into a minimum output pulse width of 3.5ns. It is foreseen to substitute soon the relatively slow chips by the DRS4 chips \cite{drs4}  that have much improved parameters.
\section{PMTS PERFORMANCES}
\begin{figure}[hbt]
\centering
\includegraphics*[scale=0.40]{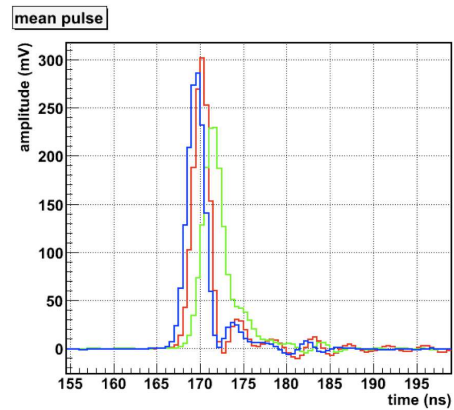}
\caption{\label{fig:pulse}
Mean pulse from different PMTs at the output of the Matacq readout system.}
\end{figure}

\begin{figure}[hbt]
\centering
\includegraphics*[scale=0.40]{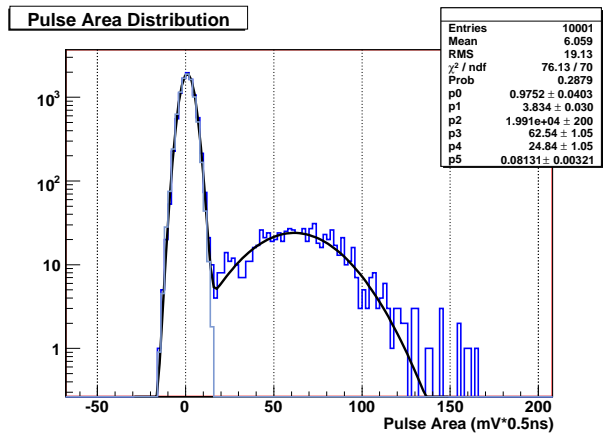}
\caption{\label{fig:spe}
The SPE distribution for 1250V. The highest peak shows the pedestal events and the second one the single photoelectron ones. The data are fitted with two Gaussian}
\end{figure}

The new PMTs have an enhanced QE with a typical peak efficiency of $\sim 32\%$.\\
\begin{figure}[hbt]
\centering
\includegraphics*[scale=0.40]{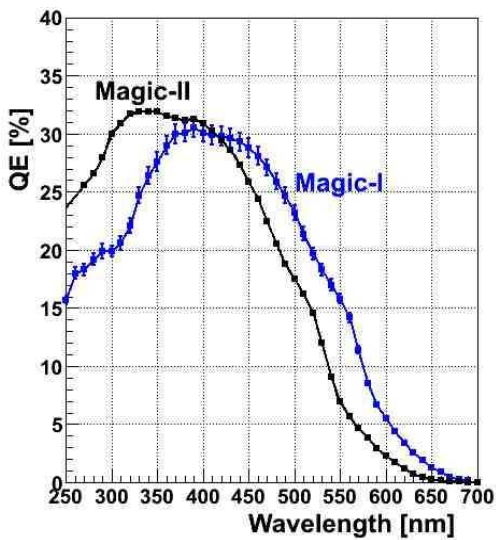}
\caption{\label{fig:qe}
QE of Magic-II PMT compared to the one of Magic-I}
\end{figure}
In Figure \ref{fig:qe} the QE curves for MAGIC-I and MAGIC-II PMTs as a function of wavelength are plotted. \\
The PMTs selected for the trigger region have a value of the cathode blue sensitivity bigger than $11.8$.\\
At the MPI in Munich, a test setup was built in order to perform some functionality and characterization tests. The test setup included a shielded box of 100cm$\times$ 60cm, in which the cluster was placed and illuminated uniformly by a laser light source. A Matacq multichannel FADC board with 300MHz bandwidth, 2GSample/s sampling and inputc range of 1V was used for the signal readout.\\
The signals measured have a pulse width of $\sim$ 2.5ns as shown in Figure \ref{fig:pulse}.\\
All the pixels tested were scanned at different voltages and then 'flatfielded' to the same gain of $3\cdot 10^4$. The result of the fits show a trend of the pulse area proportional to the voltage to the power of $\sim$ 4.
Linearity tests show that the dynamic range of the entire chain, up to the FADC used, is $\geq$ 800 photoelectrons.\\
The single photoelectron resolution (SPE) is shown in Figure \ref{fig:spe}. The measurements of the SPE are done at the highest gain for 1250V, tuning the laser light emission down to around the single photon regime. The trigger for the FADC was directly derived from the laser. The SPE pulses were extracted from FADC data and the charge was obtained by integrating each pulse over 5ns. The laser intensity was set such that the SPE event fraction was kept around 5$\%$ in order to reduce the probability of two photoelectrons coinciding within one pulse to be smaller than 0.2$\%$ of the total number of events.

\section{THE CAMERA UPGRADE WITH HPDS}
\begin{figure}[hbt]
\centering
\includegraphics*[scale=0.30]{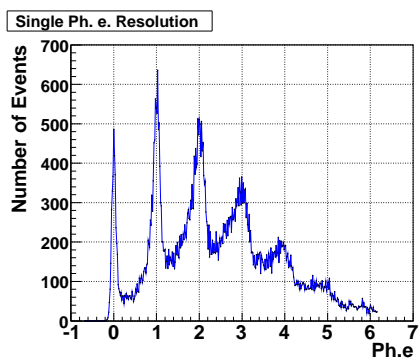}
\caption{\label{fig:hpd}
Single photoelectron resolution for HPD with a gain of $\sim 78000$}
\end{figure}

An upgrade of the camera of the MAGIC telescopes is foreseen for the near future. A new photodetector type, the Hybrid Photo Detector (HPD), will substitute part of the PMTs in the camera. HPDs consist of a vacuum tube with a GaAsP photocathode set at 6-8 kV and an avalanche diode acting as an electron bombarded anode with internal gain. Thanks to the new camera design the HPD clusters  can easily replace the PMT ones.
The good SPE resolution of HPDs and the low afterpulse rate are advantageous against PMTs.
Moreover, HPDs have a higher QE than PMTs and thus they will provide about 2 times more photoelectrons for the same input light.
Figure \ref{fig:hpd} shows an example of a photoelectron spectrum in case of a low light illuminated  HPD with a gain of $\sim 78000$; multi-photoelectrons peaks are well resolved. Further details of the tests of an HPD cluster can be found in the contribution 'Development of the HPD Cluster for MAGIC II' in this conferences.

\section{ACKNOWLEDGMENTS}
We would like to thank the technical and electronic staff of the IAC as well as of the MPI. The contribution from the German BMBF and MPG, the Italian INFN and INAF, the Spanish MCINN, the Swiss ETH and the Polish MNiSzW are vital for the MAGIC project.

\end{document}